\documentclass{PoS}
\usepackage{amssymb,amsmath,multirow,epsfig,graphicx,color}

\usepackage{epsfig}
\usepackage{graphicx,amsmath}
\usepackage{color}
\usepackage{booktabs}
\usepackage{axodraw2}
\newcommand\ba{\begin{eqnarray}}
\newcommand\ea{\end{eqnarray}}

\newcommand{\be}{\begin{equation}}
\newcommand{\ee}{\end{equation}}
\newcommand{\bas}{\begin{eqnarray*}}
\newcommand{\eas}{\end{eqnarray*}}

\newcommand {\sla} {\slash \hspace{-0.21cm}}

\def\auj{\number\day\space\ifcase\month\or
	janvier\or f\' evrier\or mars\or avril
	\or mai\or juin\orjuillet\or ao\^ut
	\or septembre\or octobre\or novembre
	\or d\' ecembre\fi\space\number\year}  
\def\hoje{\number\day\space de \ifcase\month\or
	Janeiro,\or Fevereiro,\or Mar\cc o,\or 
	Abril,\or Maio,\or Junho,\or Julho,
	\or Agosto,\or Setembro,\or Outubro,\or Novembro,
	\or Dezembro,\fi\space\number\year}
\def\data{\number\day\space  \ifcase\month\or
	January,\or February,\or March,\or 
	April,\or May,\or June,\or July,
	\or Agust,\or septembre,\or Octubre,\or November,
	\or December,\fi\space\number\year}

\title{Nucleon Electromagnetic and Axial Form Factors in a Light-front
Constituent Quark Model
\thanks{
	This work was supported in part by CAPES, and 
	Conselho Nacional de Desenvolvimento Cient\'ifico e Tecnol\'ogico (CNPq) 
	under grants 308025/2015-6 (JPBCM), 308486/2015-3 (TF).
	Funda\c{c}\~ao de Amparo \`a Pesquisa do Estado de S\~ao Paulo
	(FAPESP) under the thematic projects 2013/26258-4 and 2017/05660-0, 
	and by regular project 2019/02923-5 (JPBCM). Project INCT-FNA Proc. No. 464898/2014-5.
   }}
	
  \ShortTitle{Nucleon electromagnetic form factors}
  
  \author{\speaker{W.~R.~B.~de~Ara\'ujo  } \\  \thanks{ \bf \bf LFTC-22-3/71}
  Secretaria de Educa\c c\~ao do Estado de S\~ao Paulo, DE Norte 2, Seduc SP, Brazil
  \\  
E-mail: \email{ wilsonrbarbosa@prof.educacao.sp.gov.br}} 
  
\author{E.~F.Suisso \\
Instituto Nacional de Propriedade Industrial,INPI, RJ,
Brazil\\
    E-mail: \email{suisso@inpi.gov.br}}

\author{J. P. B. C. de Melo \\
    Laborat\'orio de F\'\i sica Te\'orica e 
     	Computacional - LFTC \\
     	Universidade Cruzeiro do Sul~/~Universidade Cidade de S\~ao Paulo, 
     	01506-000 S\~ao Paulo, Brazil  \\
     E-mail: \email{joao.mello@cruzeirodosul.edu.br}}

\author{T. Frederico \\
Instituto Tecnol\'ogico de
	Aeron\'autica, DCTA \\ 12.228-900 S\~ao Jos\'e dos Campos, SP,
	Brazil. \\ 
	E-mail: \email{tobias@ita.br}
}

\author{Kazuo Tsushima  \\
    Laborat\'orio de F\'\i sica Te\'orica e 
     	Computacional - LFTC \\
     	Universidade Cruzeiro do Sul~/~Universidade Cidade de S\~ao Paulo, 
     	01506-000 S\~ao Paulo, Brazil  \\
        E-mail: \email{kazuo.tsushima@cruzeirodosul.edu.br}}

\title{Nucleon Electromagnetic and Axial Form Factors in a Light-front
Constituent Quark Model}

\abstract{In the present work we study the effect of the scalar spin coupling 
of constituent quarks on the nucleon electroweak properties by introducing
a valence light-front wave function with two momentum scales. By comparing 
the results obtained with the one scale and  two scale wave function models, we have found that
the last one has shown a reasonable description of the static observables and 
$\mu_pG_{Ep}/G_{Mp}$ ratio in which the position of the zero appears around 10~[GeV/c]$^2$ or for
higher squared momentum transfers. We have also shown results for the
axial-vector coupling  $g_{A}$  and the nucleon axial-vector form factor. The best result
for $g_A$ was obtained when the parameters of the nucleon wave 
function model were such that the experimental 
value of the neutron magnetic moment was described.}

\FullConference{
  XV International Workshop on Hadron Physics (XV Hadron Physics)
  13 -17 September 2021 \\
  Online, hosted by Instituto Tecnológico de Aeronáutica, São José dos Campos, Brazil\\
  }

\begin{document}
\maketitle

\section{Effective Lagrangian for the N-q coupling}

The first step in our model is to wrote down the effective Lagrangian 
\begin{eqnarray}
{\cal{L}}_{N-3q}&=& m_N\epsilon^{lmn}
\overline{\Psi}_{(l)}
i\tau _2\gamma _5\Psi_{(m)}^C\overline{\Psi} _{(n)}\Psi _N+ H.C., 
\label{lag}
\end{eqnarray}
 which has been chosen to built a scalar spin coupling between a pair of quark fields $\overline{\Psi}_{(l)}$ and $\Psi_{(m)}^C$ with colorless nucleon field $\Psi _N$.
 One of the isospin matrices is $\tau_{2}$; the color indices are ${l,m,n}$; 
the totally antisymmetric tensor in color space is $\epsilon^{lmn}$. The charge conjugate field is
$\Psi ^{C}=C\overline{\Psi }^{\top}$,  where $ C=i\gamma ^{2}\gamma^{0}$ is the 
charge conjugation matrix and $m_N$ is the nucleon mass. Observe that through
Fierz transform the spin structure of the nucleon becomes quite rich, despite the 
simple structure chosen for the effective Lagrangian.

\section{Electromagnetic and Axial form factors}

 The matrix element of the plus component of the nucleon electromagnetic current ($J^+_N=J_N^0+J_N^3$),
taking into account the Drell-Yan condition with $q^+=q^0+q^3=0$, is given by 
\begin{eqnarray*}
\langle s'|J^+_N(Q^2)|s\rangle &=&\bar{u}(p',s')
\left( F_{1N}(Q^2)\gamma^++ \imath\frac{\sigma^{+\mu}Q_\mu}{2
m_N}F_{2N}(Q^2)
\right) {u}(p,s) \\
&=& \frac{p^+}{m_N}
\langle s'| F_{1N}(Q^2)-\imath\frac{F_{2N}(Q^2)}{2 m_N}
\hat{n} \cdot (\vec q_\perp \times \vec \sigma )| s \rangle \ ,
\label{jp}
\end{eqnarray*}
where $F_{1N}$ and $F_{2N}$ are the Dirac and Pauli form factors,
respectively, and $\hat{n}$ is the unit vector along the z-direction.
The electric and 
magnetic  Sachs form factors are  written in terms of Pauli and Dirac and Pauli form factors, as:
\begin{equation}
G_{EN}(Q^2)= F_{1N}(Q^2) - \frac{Q^2}{4m_N^2}F_{2N}(Q^2)
\quad  \text{and} \quad
G_{MN}(Q^2)= F_{1N}(Q^2)+F_{2N}(Q^2) \, .
\label{sach}
\end{equation}
From the above form factors we extract  the charge radius as $r^2_N 
= - 6\frac{dG_{EN}(Q^2)}{dQ^2}|_{Q^2=0}$ 
and the magnetic moment as $\mu_N= G_{MN}(Q^2=0)$.

From the plus component of the nucleon axial-vector current with the Drell-Yan condition we have that:
\begin{equation}
\langle s'|A_i^+(Q^2)|s\rangle = \bar{u}(p',s')G_A(Q^2)\gamma^+\gamma^5\frac{\tau_i}{2} u(p,s)
= \frac{p^+}{m_N}G_A(Q^2)\langle s'|\frac{\tau_i}{2}\sigma_z|s\rangle\, ,
\label{axial}
\end{equation}
with $g_A=G_A(0)$ being the axial-vector nucleon coupling.
For both calculations of the electromagnetic and axial-vector form factors, we adopted the Breit frame.
The squared momentum transfer is $Q^2 = -q^2 = q_{\perp}^2$, 
$p=(\sqrt{\frac{q^2_\perp}{4}+m^2_N},-\frac{\vec q_\perp}{2},0)$ and
$p'=(\sqrt{\frac{q^2_\perp}{4}+m^2_N},\frac{\vec q_\perp}{2},0)$ are
the nucleon momentum, before and 
after the absorption of the virtual
photon, respectively.

\begin{figure}[thb]
\begin{center}
\includegraphics[scale=.70,angle=0]{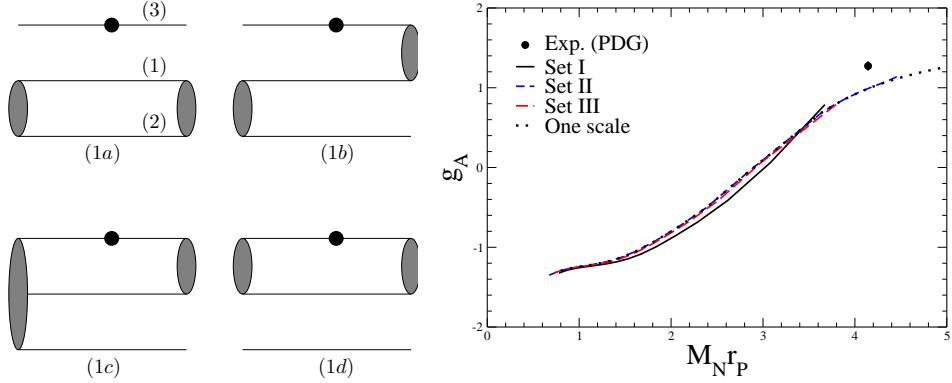}
\hspace{0.0801003cm}
\includegraphics[scale=.28,angle=0]{fig2had21.eps}
\caption{Left Panel:~Diagrammatic representation of the nucleon
photo-absorption amplitude. The gray blob represents the spin
invariant for the coupled quark pair in the effective Lagrangian
Eq.~(\ref{lag}). The small filled circle attached to the quark line represents
the action of the EM current operator.
Right panel:~Correlation between $g_A$ and $m_N r_p$ calculated by
the two-scale models (set I,  set II, Set III) and  one scale  wave function.
The experimental data from Ref.~\cite{PDG2016} is given by the filled circle.
}
\label{fig12}
\end{center}
\end{figure}

The microscopic matrix elements of the electromagnetic and axial current 
are derived from the effective Lagrangian, Eq.(\ref{lag}), within the light-front impulse 
approximation which is represented by four three-dimensional diagrams in 
Fig.\ref{fig12}~(left panel). 
The diagrams embody the antisymmetrization of the quark state in the wave function. 
In all diagrams from Fig.\ref{fig12}~(left panel), 
the quark 3 is the quark which absorbs the momentum
transfer carried by the virtual  photon.

The microscopic electromagnetic current for the nucleon depicted in 
the left panel of Fig.\ref{fig12} is given by
\begin{equation}
J^+_N(Q^2)=J^+_{aN}(Q^2)+4J^+_{bN}(Q^2)+2J^+_{cN}(Q^2)+2J^+_{dN}(Q^2)\, ,
\end{equation}
and the axial-vector current is given by, 
\begin{equation}
A_{AN}^{+}(Q^{2})=A_{aN}^{+}(Q^{2})+4A_{bN}^{+}(Q^{2})+2A_{cN}^{+}(Q^{2}) \, .
\end{equation}
For each of these microscopic currents, we performed the analytic integration over the quarks
light-front energies $k^-_i$ in the two-loop  integral.
After that, we have obtained two-loop three momentum integrations written 
in terms of the light front kinematical momenta. For instance, $J_a^+$
is given by 
\begin{eqnarray}
<s'\mid J_{aN}^+\mid s>=2 p^+ m_N^2 <N\mid\hat Q_q\mid N>
\int\frac{dk_1^+d^2\vec k_{1\perp}dk_2^+d^2\vec k_{2\perp}}
{k_1^+k_2^+k_3^{+2}}\theta(p^+-k_1^+)\theta(p^+-k_1^+-k_2^+)\nonumber\\
\times Tr\left[(\rlap\slash k_2+m)(\rlap\slash k_1+m)\right] u(p',s')(\sla k'_3+m)\gamma^+(\sla
k_3+m)u(p,s)\Psi(M'^2_0)\Psi(M^2_0)\, ,\qquad
\label{j+af}
\end{eqnarray}
where $M_0^2$ and $M_0'^2$
are the  squared free mass of the virtual  
three constituent quarks  system in the initial and final nucleon, respectively.

In the same way, we have obtained $J_{bN}^+$, $J_{cN}^+$, $J_{dN}^+$.
For the axial-vector current current, for instance $A_{aN}^+$, is given by
\begin{eqnarray}
\langle s'|A^+_{a N}(Q^2)|s\rangle =
2p^{+2}m_N^2  \langle N|\frac{
\vec{\tau}}{2}|N\rangle \int \frac{d^{2} k_{1\perp} dk^{+}_1d^{2} k_{2\perp}
d k^{+}_2 }{k^+_1k^+_2k^{+\ 2}_3} 
\theta(p^+-k^+_1)\theta(p^+-k^+_1-k^+_2)\nonumber \\
\times{\mathrm{Tr}} \left[ (\rlap\slash k_2+m) (\rlap\slash k_1+m)\right]
\bar u(p',s')(\rlap\slash k'_3+m))\gamma^+\gamma^5(\rlap\slash
k_3+m)u(p,s) \Psi (M^{'2}_0) \Psi (M^2_0)\, .
\label{A+alf}
\end{eqnarray}

Following the same 
 steps for the nucleon axial-vector current $A^+_{a N}(Q^2)$, 
 after integration over $k^-_i$ we obtain $A^+_{b N}(Q^2)$, $A^+_{c N}(Q^2)$, 
 and $A^+_{d N}(Q^2)=0$ due to the isoscalar nature of the scalar quark pair.
 
\section{Two scale wave function}

The ``two-scale model'' of the momentum component of the valence 
light-front 
nucleon wave function~\cite{Araujo2006,NPA2018} was chosen to be represented by 
a power law form~\cite{bsch,brodsky} including  two terms
\begin{equation}
\Psi_{\mathrm{Power}} = N_{\mathrm{Power}}\left[(1+M^2_0/\beta^2)^{-p}
+\lambda (1+M^2_0/\beta_1^2)^{-p_1} \right]\ ,
\label{wf1}
\end{equation}
where
\begin{equation}
\lambda = (1 + M_H^2/\beta_1^2)^{p_1}/(1 + M_H^2/\beta^2)^{p}\, .
\label{lambda}
\end{equation}
The characteristic momentum scales of the wave
function \eqref{wf1} are represented by~$\beta$,~$\beta_1$~and~$M_H$, 
where the low-momentum scale, $\beta\sim \Lambda_{QCD}$ is driven essentially
by the nucleon static observables, while the high-momentum scales 
are driven 
by the zero of $G_{Ep}(Q^2)$~\cite{Araujo2006,NPA2018}. 
The mass scale $M_H$ gives the mass of the virtual three-quark system where the
two terms in the wave function have the same magnitude. We associate this mass scale as a
qualitative measure of the boundary separating the IR dynamics  and 
the one-gluon exchange interaction dominance. Note that our choice corresponds to a node-less momentum
component of the valence wave function in order to represent the nucleon ground state.
The normalization constant $N_{\mathrm{Power}}$ is
fixed to reproduce the  proton charge. 

\begin{table}[htb]
\caption{Parameters of one and two scales wave functions}
\label{table2}
\begin{center}
\begin{tabular}{|l|c|c|c|c|c|c|}
\hline
\hline
Reference    & $\beta$~(GeV) & $\beta_1$~(GeV) &$p$ & $p_1$  & $M_{H}$~(GeV)
\\
\hline
Set I     & 1.07   & 10    & 3.4& 3 & 3.68\\
Set II    & 0.396  & 10.56 & 3  & 3 &5.92   \\
Set III   & 0.34   & 7.5   & 3.2& 3 & 4.32\\
One scale I & 0.477   &  ~~-& 3  & ~~-&~~-\\
One scale II & 1.07   &  ~~-& 3  & ~~-&~~-\\
\hline
\hline
\end{tabular}
\end{center}
\label{table 1}
\end{table}

\section{Numerical Results}

In  our models we have used the constituent quark mass $m$ of 220 MeV. In Table\ref{table 1}, we show
the parameters Set I and one scale II, chosen to reproduce
the proton magnetic moment, $\mu_p$, respectively, and  Set II, Set III and 
one scale II, chosen to reproduce the neutron magnetic moment, $\mu_n$, respectively.
The static observables are preserved by introducing the second scale in the 
wave function. This can be checked by
looking to Table~\ref{table 2}.
 
\begin{figure}[htb]
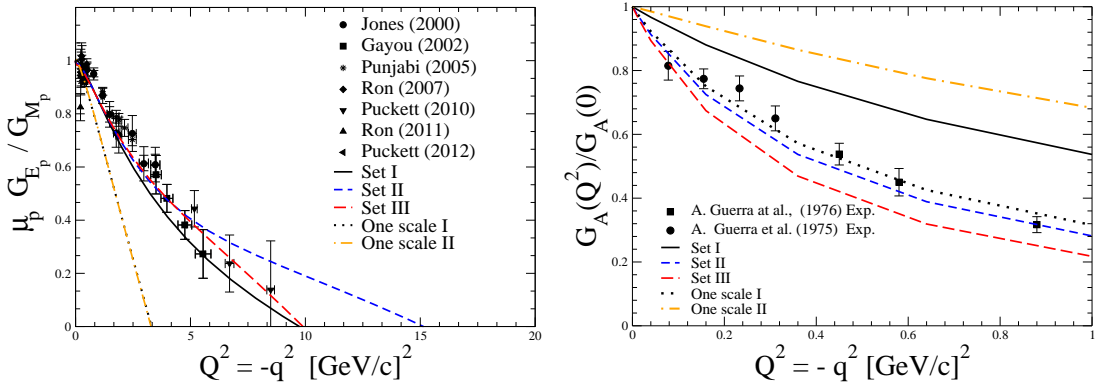

\begin{center}
\includegraphics[scale=.28,angle=0]{fig3had21.eps}
\hspace{0.0801003cm}
\includegraphics[scale=.28,angle=0]{fig4had21.eps}
\caption{Left Panel:
~Proton electromagnetic form factor ratio $\mu_p G_{Ep}(Q^2)/G_{Mp}(Q^2)$,
calculated by the two-scale  (set I, set II, Set III)
and one-scale I and II  wave functions models. The experimental data are from
Refs.~\cite{Jones2000,Brash2002,Gayou2002,Punjabi2005,Ron2007,Puckett2010,
Ron2011,Puckett2012}.
Right panel:~Normalized axial-vector form factor, $G_A(Q^2)/G_A(0)$,
calculated with the two-scale (set I, set II and set III), and one-scale (I and II)
wave functions. Experimental data are from~\cite{Guerra}.
}
\label{fig34}
\end{center}
\end{figure}

\begin{table}[htb]
\caption{Axial and Electromagnetic properties of nucleons, and the value of zero
$Q_0^2$ of $G_{Ep}(Q_0^2)=0$ obtained with the two-scale (set I and set II)
and one-scale wave functions.}
\vspace{0.25cm}
\label{table1}
\begin{center}
\begin{tabular}{|l|c|c|c|c|c|c|c|}
\hline
\hline
 &$g_A$& $\mu_p~(\mu_N)$ &~$r_p$~(fm) & $\mu_n~(\mu_N)$&
$r^2_n$~(fm$^2)$  & $Q_0^2$~(GeV$^2$) \\
\hline
Set I & 0.73   & 2.80 & 0.78 & -1.52  & -0.07 &  8.27  \\
Set II & 1.10    & 3.05   & ~0.94  & -1.88  & -0.06 &  15.12  \\
Set III &  1.17    & 3.11 & 1.00   & -1.97  & -1.01 &  9.94  \\
One scale 1 & 1.01 & 3.11   & ~1.03  & -1.91  & -0.08 &  3.28   \\
One scale 2& 0.52  & 2.79   & ~0.75  & -1.51  & -0.10   &  3.28   \\
\hline
\hline
\end{tabular}
\label{table 2}
\end{center}
\end{table}

The left panel in Fig.\ref{fig34} shows the $\mu_pG_{E_p}/G_{M_p}$ ratio as a function of the square momentum transfer for all the five models given in Table~\ref{table 1}. The two scale models present the zero of the proton electric form factor for $Q^2_0\sim 10$~ GeV$^ 2$, showing a dramatic improvement over the one scale models in comparison to the experimental data. The introduction of the
high momentum scale increases the   probability of  the three-quark system to be found with high virtuality, associated with the long  tail of the momentum component of the wave function.

Our model shows a better description of the experimental data
when the parameters are chosen to fit neutron magnetic moment.
We have observed this behaviour for all nucleon electromagnetic and axial-vector  form factors.
For instance, the pattern shown in the right panel of Fig.\ref{fig34},  
was also found for the proton and neutron electromagnetic form factors.

\section{Conclusion and Summary}

We give a brief conclusion and summary from our work:
\begin{enumerate} 
\item The two-scale nucleon wave function model carrying a
 high-momentum scale, can achieve a reasonable description of the ratio
$\mu_p G_{Ep}(Q^2) / G_{Mp}(Q^2)$ and the position of its zero,
without destroying the good outcomes provided by  the one-scale wave function model
in describing the nucleon static properties.

\item The best description of $\mu_p G_{Ep}/G_{Mp}$ and static observables
is achieved by the introduction of the two-scale wave function with the parameters
 chosen to reproduce  $\mu_n$.

\item A good description of the nucleon electroweak observables, namely the axial-vector and electromagnetic form factors, is found when the wave function  parameters are
chosen to reproduce the neutron magnetic moment.
\end{enumerate}

{\it \bf  Acknowledgements:}~
This work was supported in part by CAPES under Grant 
and
by the Conselho Nacional de Desenvolvimento 
Cient\'{i}fico e Tecnol\'{o}gico (CNPq),
~Grant No.~308486/2015-3 (TF),
Process No.~307131/2020-3 (JPBCM),
Grants No. 438562/2018-6 and 
Funda\c{c}\~{a}o de Amparo \`{a} Pesquisa do Estado de S\~{a}o Paulo (FAPESP), 
Process No. 2019/02923-5 (JPBCM),  
and was also part of the projects, Instituto Nacional de Ci\^{e}ncia e
Tecnologia -- Nuclear Physics and Applications (INCT-FNA), Brazil,
Process No.~464898/2014-5, and FAPESP Tem\'{a}tico, Brazil, Process,
the thematic projects, No. 2013/26258-4 and No. 2017/05660-0.
KT was supported by
the Conselho Nacional de Desenvolvimento Cient\'\i fico e Tecnol\'ogico (CNPq) Process, No. 313063/2018-4, and No. 426150/2018-0, and Funda\c c\~ ao de amparo \`a Pesquisa do Estado de S\~ao Paulo (FAPESP) Process, No. 2019/00763-0.

\end{document}